\documentclass[aps,pra,10pt]{revtex4-2}

\usepackage{amsfonts}
\usepackage{graphicx}
\usepackage{amssymb}
\usepackage{amsmath}
\usepackage{rotating}

\begin{document}

\title{Effect of impurity in multi-species medium-body boson systems at ultracold temperatures}

\author{Chengguang Bao}\thanks{Corresponding author: C. G. Bao, stsbcg@mail.sysu.edu.cn}
\author{Yanzhang He}

\affiliation{School of Physics, Sun Yat-Sen University, Guangzhou, 510275, P. R. China}

\date{\today}

\begin{abstract}
An approach is proposed for exactly solving multi-species medium-body
systems with spin-1 atoms at ultracold temperature. The ground state (g.s.)
and the low-lying excitation modes have been studied in detail. Both exact
numerical solutions and approximate analytical solutions have been obtained.
With the latter, an analytical investigation is performed to unveil the
underlying physics. The competition between the repulsive intra-species and
the inter-species interactions is found to be essential. The former
interaction leads to the formation of 0-pairs (two spins are coupled to
zero) while the latter leads to the formation of polarized clusters (a
subsystem in which all the spins of a kind of atoms point to a common
direction). This competition is distinctive in multi-species systems; it
governs the textures and variations of the spins. Another distinctive
feature is the emergence of the odd-channel for two different bosons. The
effect of the odd-channel is found to be important when the particle numbers
are large. Thus, for the substances with the strength of the odd-channel
being not so weak, the previous many-body theories should be revised if it
has not taken this channel into account. A great effort is placed on
clarifying the effect of impurity. A number of formulae have been obtained
to guide the change of magnetic properties of a material by introducing
appropriate impurities. In particular, we find that the change against the
increase of the particle number of an impurity might not be smooth. There
are critical values, when the number crosses over a critical value, a great
change of the property can suddenly occur. This reveals a new pathway for
inducing quantum phase transitions.
\end{abstract}

\maketitle

\section{Introduction}

An important technological achievement is the realization of extremely low
temperatures, e.g., $T\leq 100~\text{pK}$ \cite{cdprl21}. This makes the
field of ultracold physics increasingly important. When $T$ is close to
zero, the spin-degrees of freedom become dominant while the spatial degrees
of freedom are nearly frozen. This paper is limited to ultracold physics
with the following aims:

(i) The first aim is to propose an approach for solving medium-body cold
systems to obtain exact numerical solutions together with approximate
analytical solutions. In this way, the theory might serve as a bridge to
connect many-body physics and few-body physics. In particular, our exact
solutions allow for a stringent test of many-body approximations and clarify
which few-body phenomena would persist in accord with the increase of
particle numbers. Our analytical solutions allow for revealing the
underlying physics so that how various physical factors are competing could
be understood.

(ii) The second aim is to study multi-species ultracold condensates. There
are a lot of papers dedicated to the study of single-species condensates,
\cite{ref5,stent98} but fewer for multi-species condensates. \cite{hos,esry97,pu98,man,mlpra07,ref14} However, the latter have two
distinguished physical factors, namely, the competition between the intra-
and inter-species interactions and the appearance of the odd channel, where
two different bosons are coupled to $\lambda\hbar$ with $\lambda$ being an
odd integer. Accordingly, the physics involved in a multi-species system is
remarkably richer.

In particular, this study provides a rigorous approach for studying the
effect of impurity. This approach is valuable not only in the academic
aspect but also in the application aspect (say, in the preparation of
multi-species artificial cold systems with a desired property).

(iii) The third aim is to study various low-lying excitation modes. In
particular, the possible emergence of low-lying collective mode in
multi-species systems is notable. This study generalizes the study on the
ground state (g.s.) and is crucial for understanding the transitions between
various spin-textures. \cite{ref14,hlb25},

(iv) The fourth aim is to clarify the effect of the odd-channel. Since this
channel might play an important role as shown in this paper, The study would
be necessary for the preparation of artificial cold systems.

\section{Spin-dependent Hamiltonian of cold systems with two kinds of spin-1 atoms}

\subsection{The Hamiltonian with three parameters}

We consider a mixture of spin-1 atoms, $N_{A}$ $A$-atoms and $N_{B}$ $B$
-atoms, bound in an optical trap. The particle numbers are assumed to be
even and $N_{A}\geq N_{B}$, for convenience. The temperature is assumed to
be sufficiently low so that the spatial degrees of freedom are frozen.
Accordingly, the spatial wave function for every $A$-atom ($B$-atom) is
fixed at $\Phi _{A}$ ($\Phi _{B}$). Only the spin degrees of freedom are
considered as active and the spin textures depend primarily on the
spin-dependent force, while the central force can be neglected. Furthermore,
let $\lambda $ be the combined spin of two atoms. Obviously, when the two
atoms belong to the same species, $\lambda =0$ or $2$; otherwise, the
additional $\lambda =1$ channel emerges. When we carry out an integration
over all the spatial degrees of freedom, we obtain the spin-dependent
Hamiltonian
\begin{equation}
\hat{H}_{\text{spin}}\ =\sum_{\lambda ,i>j}\tilde{g}_{\lambda
}^{A}P_{\lambda }^{A,i,j}+\sum_{\lambda ,i>j}\tilde{g}_{\lambda
}^{B}P_{\lambda }^{B,i,j}+\sum_{\lambda ,i,i^{\prime }}\tilde{g}_{\lambda
}^{AB}P_{\lambda }^{AB,i,i^{\prime }}\equiv \hat{H}_{A}+\hat{H}_{B}+\hat{H}
_{AB}.  \label{h}
\end{equation}
$P_{\lambda }^{X,i,j}$ ($X=A$ or $B$) is the projector for extracting the
component where the two $X$-atoms $i$ and $j$ are coupled to $\lambda =0$ or
$2$. $\tilde{g}_{\lambda }^{X}\equiv g_{\lambda }^{X}\int \Phi _{X}^{4}
\mathrm{d}r$, $g_{\lambda }^{X}$ is the strength of the intra-species
interaction of the $X$ -species related directly to the phase shift of the $
\lambda $-channel. $P_{\lambda }^{AB,i,j}$ is the projector for extracting
the component where the $i$-th $A$-atom and the $j$-th $B$-atom are coupled
to $\lambda =0$, $1$, or $2$, $\tilde{g}_{\lambda }^{AB}=g_{\lambda
}^{AB}\int \Phi _{A}^{2}\Phi _{B}^{2}\mathrm{d}r$, and $g_{\lambda }^{AB}$
is the strength of the inter-species interaction in the $\lambda $-channel.
The total spin $S$ of the binary system is conserved. However, due to the
effect of the $\lambda =1$ channel, the two combined spins of the two
species denoted as $S_{A}$ and $S_{B}$, respectively, are no more conserved.

Note that there are totally seven parameters $\tilde{g}_{\lambda }^{Z}$ ($
Z=A $, $B$, or $AB$) contained in $\hat{H}_{\text{spin}}$. To facilitate the
related analysis, we are going to identify a much smaller parameter space
without loss of physics.

Let $\hat{\mathbf{s}}_{i}$ be the spin operator and $\chi (i)$ be the
spin-state of the $i$-th atom.\ When the spins of the $i$-th and $j$-th
atoms are coupled to $\lambda $ denoted as $(\chi (i)\chi (j))_{\lambda }$,
we have
\begin{equation}
\hat{\mathbf{s}}_{i}\cdot \hat{\mathbf{s}}_{j}(\chi (i)\chi (j))_{\lambda
}=q_{\lambda }(\chi (i)\chi (j))_{\lambda },  \label{ssq}
\end{equation}
where $(\chi (i)\chi (j))_{\lambda }$ is called a $\lambda $-$Z$-pair and $
q_{\lambda }=-2$, $-1$, and $1$ for $\lambda =0$, $1$, and $2$,
respectively, where the $\lambda =1$ channel exists only if $Z=AB$. From
Eq.~(\ref{ssq}) and the basic feature of projectors $\sum_{\lambda
}P_{\lambda }^{Z,i,i^{\prime }}=1$, we can establish a relation between the
spin operators and the projectors. It appears as $P_{0}^{Z,i,j}=\frac{1}{3}
(1-\hat{\mathbf{s}}_{i}\cdot \hat{\mathbf{s}}_{j})-2P_{1}^{Z,i,j}$ and $
P_{2}^{Z,i,j}=\frac{1}{3}(2+\hat{\mathbf{s}}_{i}\cdot \hat{\mathbf{s}}
_{j})-P_{1}^{Z,i,j}$ (where $P_{1}^{Z,i,j}$ appears only if $Z=AB$). Then,
we found
\begin{eqnarray}
\hat{H}_{A} &=&a\hat{S}_{A}^{2}+C_{A},  \label{ha} \\
\hat{H}_{B} &=&b\hat{S}_{B}^{2}+C_{B},  \label{hb} \\
\hat{H}_{AB} &=&2c\hat{S}_{A}\cdot \hat{S}_{B}+C_{AB}+\sum_{i,i^{\prime
}}dP_{1}^{AB,i,i^{\prime }},  \label{hab}
\end{eqnarray}
where $\hat{S}_{X}=\sum_{i\in X}\hat{\mathbf{s}}_{i}$ is the
operator for $S_{X}$, $a=\frac{1}{6}(\tilde{g}_{2}^{A}-\tilde{g}_{0}^{A})$, $
b=\frac{1}{6}(\tilde{g}_{2}^{B}-\tilde{g}_{0}^{B})$, $c=\frac{1}{6}(\tilde{g}
_{2}^{AB}-\tilde{g}_{0}^{AB})$, and $d=\frac{1}{3}(3\tilde{g}_{1}^{AB}-
\tilde{g}_{2}^{AB}-2\tilde{g}_{0}^{AB})$.

The three $C_{Z}$ are constants, they would only shift the spectra as a
whole but do not affect the details of the spin-states, therefore they can
be dropped. Thus, instead of seven, all physics remains in the 4-dimensional
parameter space expanded by $a$, $b$, $c$, and $d$. The implications of
these parameters are as follows. When $a$ is negative, i.e., $\tilde{g}
_{0}^{A}>\tilde{g}_{2}^{A}$, every pair of atoms would prefer to form the 2-$
A$-pairs (so as to lower the energy), two or more 2-$A$-pairs would also
prefer to have their spins pointing at a common direction to form a fully
polarized cluster (this subsystem is called in the f-phase) to maximize $
S_{A}$ so as to minimize $\hat{H}_{A}$ (as shown in Eq.(\ref{ha})). Whereas when $a$ is positive, the atoms would prefer
to be two-by-two to form the 0-$A$-pairs (it is called in the p-phase) to
minimize $S_{A}$ so as to minimize $\hat{H}_{A}$. Thus $a$ measures the
ability to keep the $A$-species in f-phase (if $a<0$) or in p-phase (if $a>0$
). Similarly, $b$ measures the ability to keep the $B$-species in f-phase
(if $b<0$) or in p-phase (if $b>0$). Besides, a negative $c$ would help to
maximize both $S_{A}$ and $S_{B}$\ and make them lie along the same
direction (as shown in Eq.(\ref{hab})), while a
positive $c$ would also help to maximize $S_{A}$ and $S_{B}$ but push them
to lie along opposite directions. Thus, $c$ affects not only the magnitudes
of $\mathbf{S}_{A}$ and $\mathbf{S}_{B}$ but also their relative
orientation. For example, when both $a>0$, $b>0$, and $c$ is weak, both
species would prefer the p-phase. However, when $c$ is sufficiently negative
(positive), there is a competitive choice: both species change to the
f-phase with their spins parallel (antiparallel) so as to maximize the
effect of $c$. Finally, the phase of the g.s. (f-f or p-p) depends on
competition between the intra- and inter-species interactions.

Furthermore, since the eigenstates do not depend on which energy unit is
used, when the norm of $c$ is used as the unit of energy, Eq.~(\ref{h})
becomes
\begin{eqnarray}
\hat{H}_{\text{spin}}/|c| &=&\hat{H}_{0}+\hat{H}_{1},  \label{hc} \\
\hat{H}_{0} &=&(a^{\prime }\pm 1)\hat{S}_{A}^{2}+(b^{\prime }\pm 1)\hat{S}
_{B}^{2}\mp \hat{S}^{2},  \label{hc0} \\
\hat{H}_{1} &=&\sum_{i,i^{\prime }}d^{\prime }P_{1}^{AB,i,i^{\prime }},
\label{hc1}
\end{eqnarray}
where $a^{\prime }\equiv a/|c|$, $b^{\prime }\equiv b/|c|$, and $d^{\prime
}\equiv d/|c|$. The upper (lower) signs of $\pm $ and $\mp $ are used for $
c<0$ ($c>0$) and the square of the total spin $\hat{S}^{2}=(\hat{S}_{A}+\hat{
S}_{B})^{2}$.

Thus, instead of seven, only the three parameters $a^{\prime }$, $b^{\prime
} $, and $d^{\prime }$ are sufficient to describe all the physics.

We are going to find out the exact eigenstates of $\hat{H}_{\text{spin}}/|c|$. This can be achieved via a numerical diagonalization against a set of
complete basis-states. Note that, for the $X$-species alone, the combined
spin $S_{X}$ is a good quantum number. The physical acceptable values of $
S_{X}$ should lie between $0$ and $N_{X}$, and has the same even-odd parity
as $N_{X}$ has. Let the associated normalized and symmetrized eigenstate be
denoted as $\theta _{S_{X}}^{N_{X}}$ which is a mixture of an $X$-cluster
(where a group of $X$-atoms have their spins pointing at a common direction
and form a polarized cluster with a total spin $S_{X}$) together with $\frac{
1}{2}(N_{X}-S_{X})$ 0-$X$-pairs. If $S_{X}=N_{X}$, the eigenstate is in
f-phase, if $S_{X}=0$, the eigenstate is in p-phase as mentioned. If $
0<S_{X}<N_{X}$, the state is called in q-phase. The set $\{\theta
_{S_{X}}^{N_{X}}\}$ is complete. \cite{ref10} Therefore the set $\{(\theta
_{S_{A}}^{N_{A}}\theta _{S_{B}}^{N_{B}})_{S}\}\equiv \{\phi _{S_{A}S_{B}S}\}$
is complete for the binary spin-system, where $S_{A}$ and $S_{B}$ are
coupled to $S$ ranging from $|S_{A}-S_{B}|$ to $S_{A}+S_{B}$.

When the set $\{\phi_{S_AS_BS}\}$ are used as basis-states, we have to deal
with the complicated $\theta_{S_X}^{N_X}$. However, due to the introduction
of the fractional parentage coefficients \cite{fpc} given in the Appendix,
all related 1-body and 2-body matrix elements can be easily obtained without
using the explicit expression of $\theta_{S_X}^{N_X}$. In particular, we
obtain
\begin{eqnarray}
\langle \phi_{S_A^{\prime }S_B^{\prime }S}| \frac{1}{|c|} \hat{H}_\text{spin}
| \phi_{S_AS_BS} \rangle &=& \delta_{S_A^{\prime }S_A} \delta_{S_B^{\prime
}S_B} [ (a^{\prime }\pm 1) S_A(S_A+1) +(b^{\prime }\pm 1) S_B(S_B+1) \mp
S(S+1) ] +N_A N_B \\
&&\times \sum_{L,J_A,J_B,I_A,I_B} d^{\prime }\mathfrak{U}(1,LST_A^{
\prime(J_A)}T_B^{^{\prime }(J_B)}) \mathfrak{U}(1,LST_A^{(I_A)}T_B^{(I_B)})
\delta_{T_A^{\prime(J_A)},T_A^{(I_A)}}
\delta_{T_B^{\prime(J_B)},T_B^{(I_B)}},  \nonumber  \label{hm}
\end{eqnarray}
where, in the last term, $L$ runs from $|1-S|$ to $1+S$, and all the four
indexes $J_A$, $J_B$, $I_A$, $I_B$, run from $1$ to $2$, $T_X^{(1)}=S_X+1$, $
T_X^{(2)}=S_X-1$, and $T_X^{^{\prime }(1)}=S_X^{\prime }+1$, $T_X^{^{\prime
}(2)}=S_X^{\prime }-1$,
\begin{eqnarray}
\mathfrak{U}(1,LST_A^{(I_A)}T_B^{(I_B)}) = X_A^{(I_A)} X_B^{(I_B)} \sqrt{
3(2L+1)(2S_A+1)(2S_B+1)} U \left\{
\begin{array}{ccc}
1 & 1 & 1 \\
T_A^{(I_A)} & T_B^{(I_B)} & L \\
S_A & S_B & S
\end{array}
\right\},
\end{eqnarray}
where the label $U$ with the nine indexes is the 9-$j$ coefficient for the
spins re-coupling, and $X_X^{(1)}=a_{S_X}^{N_X}$, $X_X^{(2)}=b_{S_X}^{N_X}$,
refer to Eq.~(\ref{fpc1}).

With the set $\{\phi _{S_{A}S_{B}S}\}$ and the above formulae, numerical
diagonalization of the Hamiltonian can be realized to obtain exact
solutions. However, before presenting the exact results, we will make a
qualitative analysis based on neglecting the odd $\lambda $=1 channel.

\subsection{The eigenstates of $\hat{H}_0$, the phase diagram and the excitation of the g.s. ($c<0$)}

When the odd-channel is neglected, it is clear from Eq.~(\ref{hc0}) that $
\phi _{S_{A}S_{B}S}\equiv (S_{A},S_{B})_{S}$ is an eigenstate of $\hat{H}
_{0} $ with the eigenenergy
\begin{equation}
E_{S_{A}S_{B}S}=(a^{\prime }\pm 1)S_{A}(S_{A}+1)+(b^{\prime }\pm
1)S_{B}(S_{B}+1)\mp S(S+1),  \label{hc00}
\end{equation}
where the upper (lower) signs of $\pm $ and $\mp $ are for $c<0$ ($c>0$).
Obviously, regardless of how the values of $S_{A}$ and $S_{B}$ would be, the
g.s. will have $S=S_{A}+S_{B}$ ($|S_{A}-S_{B}|$) for $c<0$ ($>0$) to
minimize the energy.

In the following, a detailed study of the case with $c<0$ is given. The case
with $c>0$ is similar and is only given by a short discussion. For
simplicity, we introduce $E_{S_{A}S_{B},+}\equiv E_{S_{A}S_{B},S_{A}+S_{B}}$
and $(S_{A},S_{B})_{+}\equiv (S_{A},S_{B})_{S_{A}+S_{B}}\equiv \phi
_{S_{A}S_{B},S_{A}+S_{B}}$.
\begin{equation}
E_{S_{A}S_{B},+}=a^{\prime }S_{A}(S_{A}+1)+b^{\prime
}S_{B}(S_{B}+1)-2S_{A}S_{B},\ (\text{if }c<0).  \label{hc01}
\end{equation}
Once the parameters $a^{\prime }$ and $b^{\prime }$ are given, from the set $
\{E_{S_{A}S_{B},S}\}$ given in Eq.~(\ref{hc00}), it is straightforward to
find out the good quantum numbers for the g.s. $\phi _{\text{gs}}$ and the
first excited state $\phi _{\text{ex}1}$, together with the lowest
excitation energy (the gap) $E_{\text{gap}}\equiv E_{\text{ex}1}-E_{\text{gs}
}$. In order to reveal the underlying physics, we first consider the
potential modes for the lowest excitation.

\subsubsection{Candidates of the lowest excitation modes ($c<0$)}

There are the following competing lower modes

(i) An increase of $S_{A}$ by two, i.e., a transition of a 0-$A$-pair to a 2-
$A$-pair (denoted as $(0\rightarrow 2)_{A}$ transition), so that $
(S_{A},S_{B})_{+}$ transits to $(S_{A}+2,S_{B})_{+}$. The associated
excitation energy
\begin{equation}
E_{\text{x}S_{A}+}=a^{\prime }(4S_{A}+6)-4S_{B}.  \label{exap}
\end{equation}
Thus, a strong $B$-cluster (i.e., a larger $S_{B}$) would help the $
(0\rightarrow 2)_{A}$ transition and thereby aid the formation of the 2-$A$
-pairs to form a polarized $A$-cluster, while a strong $a^{\prime }>0$ would
hinder the $(0\rightarrow 2)_{A}$ transition.

(ii) A decrease of $S_{A}$ by two, i.e., a transition from $
(S_{A},S_{B})_{+} $ to $(S_{A}-2,S_{B})_{+}$ and accordingly a $
(2\rightarrow 0)_{A}$ transition occurs. The associated excitation energy
\begin{equation}
E_{\text{x}S_{A}-}=a^{\prime }(-4S_{A}+2)+4S_{B}.  \label{exan}
\end{equation}
Thus a strong $B$-cluster would hinder the formation of the 0-$A$-pairs,
while a strong $a^{\prime }>0$ favors the $(2\rightarrow 0)_{A}$ transition.

(iii) An increase of $S_B$ by two, i.e., a $(0\rightarrow 2)_B$ transition
occurs. The associated excitation energy
\begin{equation}
E_{\text{x}S_B+} = b^{\prime }(4S_B+6) -4S_A
\end{equation}
In this case a polarized $A$-cluster favors the $(0\rightarrow 2)_B$
transition while a strong $b^{\prime }>0$ hinders it.

(iv) A decrease of $S_{B}$ by two, i.e., a $(2\rightarrow 0)_{B}$ transition
occurs. The associated excitation energy
\begin{equation}
E_{\text{x}S_{B}-}=b^{\prime }(-4S_{B}+2)+4S_{A}.  \label{exbn}
\end{equation}
In the above modes (i) to (iv), only a pair of particles are involved.

(v) Both $S_{A}$ and $S_{B}$ remain unchanged but a decrease of $S$ by one,
i.e., a transition from $(S_{A},S_{B})_{+}$ to $
(S_{A},S_{B})_{S_{A}+S_{B}-1} $. The associated excitation energy is
\begin{equation}
E_{\text{x}S-}=2(S_{A}+S_{B}).  \label{exsn}
\end{equation}
Recall that the unit for energy is $|c|$, thus this mode depends only on the
inter-species interaction. In fact, in this mode, both internal spin
textures remain unchanged, but the relative orientation of their
polarizations changes.

(vi) The modes in which two or a few pairs are involved. For an example, the
mode with the simultaneous formation of a 0-$A$-pair together with a 0-$B$
-pair with the excitation energy $E_{\text{x},S_{A}-,S_{B}-}$.

(vii) Collective modes in which numerous particles are involved. This mode
might emerge when the parameters are given close to the boundary separating
the f-f and p-p phases (refer to the phase-diagram).

In a specific case, the actual lowest excitation arises from the winner of
the above competitive modes.

\subsubsection{The ground state and its excitation}

\begin{figure}[tbp]
\centering\resizebox{0.9\columnwidth}{!}{\includegraphics{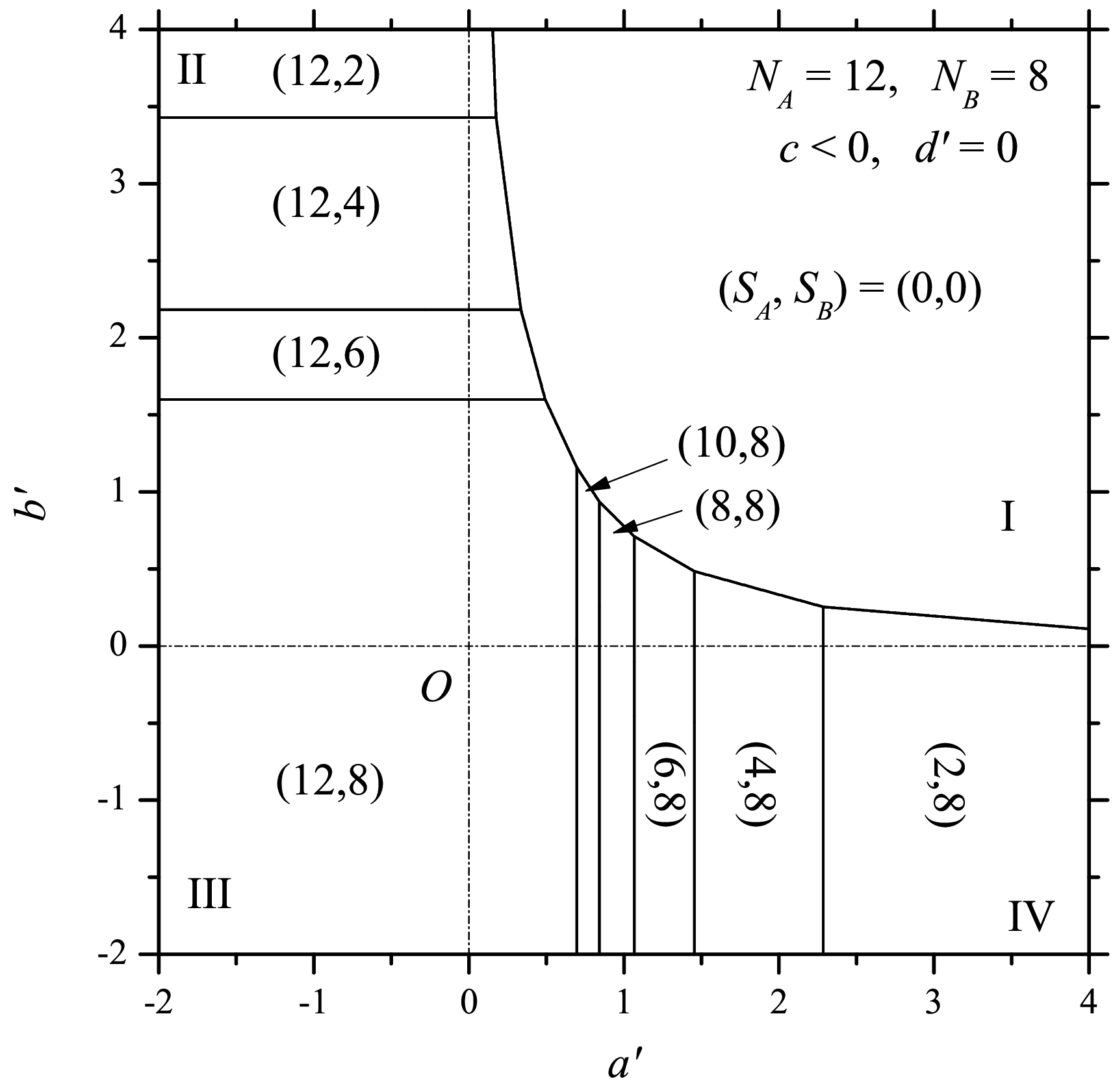}}
\caption{ The phase diagram of the g.s. of $\hat{H}_0$ against $a^{\prime }$
and $b^{\prime }$. $c<0$, $N_A=12$, $N_B=8$ are assumed. The phase
associated with a domain is specified by $(S_A,S_B)$, and the total spin $S=S_A+S_B$ }
\label{fig1}
\end{figure}

Let $N_{A}=12$, $N_{B}=8$ be assumed. With the good quantum numbers of gs, the
phase-diagram of the g.s. against $a^{\prime }$ and $b^{\prime }$ is plotted
in Fig.~\ref{fig1} where $c<0$ is assumed.

\begin{figure}[tbp]
\centering\resizebox{0.9\columnwidth}{!}{\includegraphics{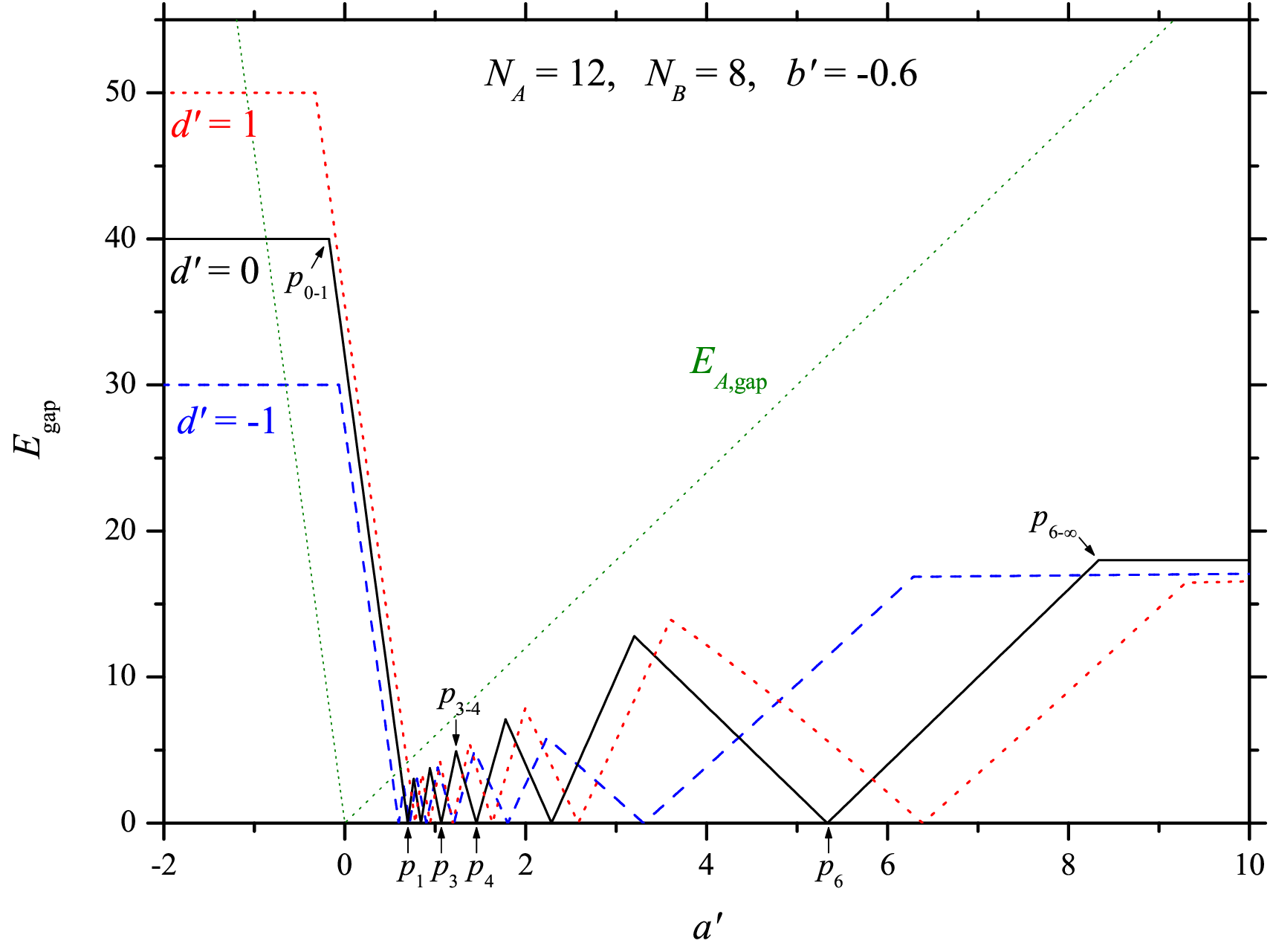}}
\caption{ (color online) $E_\text{gap}$ against $a^{\prime }$ with $
b^{\prime }=-0.6$, $N_A=12$, $N_B=8$, and $c<0$. $d^{\prime }$ has been
given in three values as $1$, $0$, and $-1$, marked by the associated curve.
Beside, $E_{A,\text{gap}}$ is also plotted in dotted broken line (for the
case with $N_B=0$). The unit for energy is $|c|$ which can be known from the
identity $|c|\equiv |b/b^{\prime }|=1.667|b|$ when $b$ has been known. There
are a number of critical points marked as $p_i$, $p_{i-(i+1)}$, $\cdots$. }
\label{fig2}
\end{figure}

\begin{figure}[tbp]
\centering\resizebox{0.9\columnwidth}{!}{\includegraphics{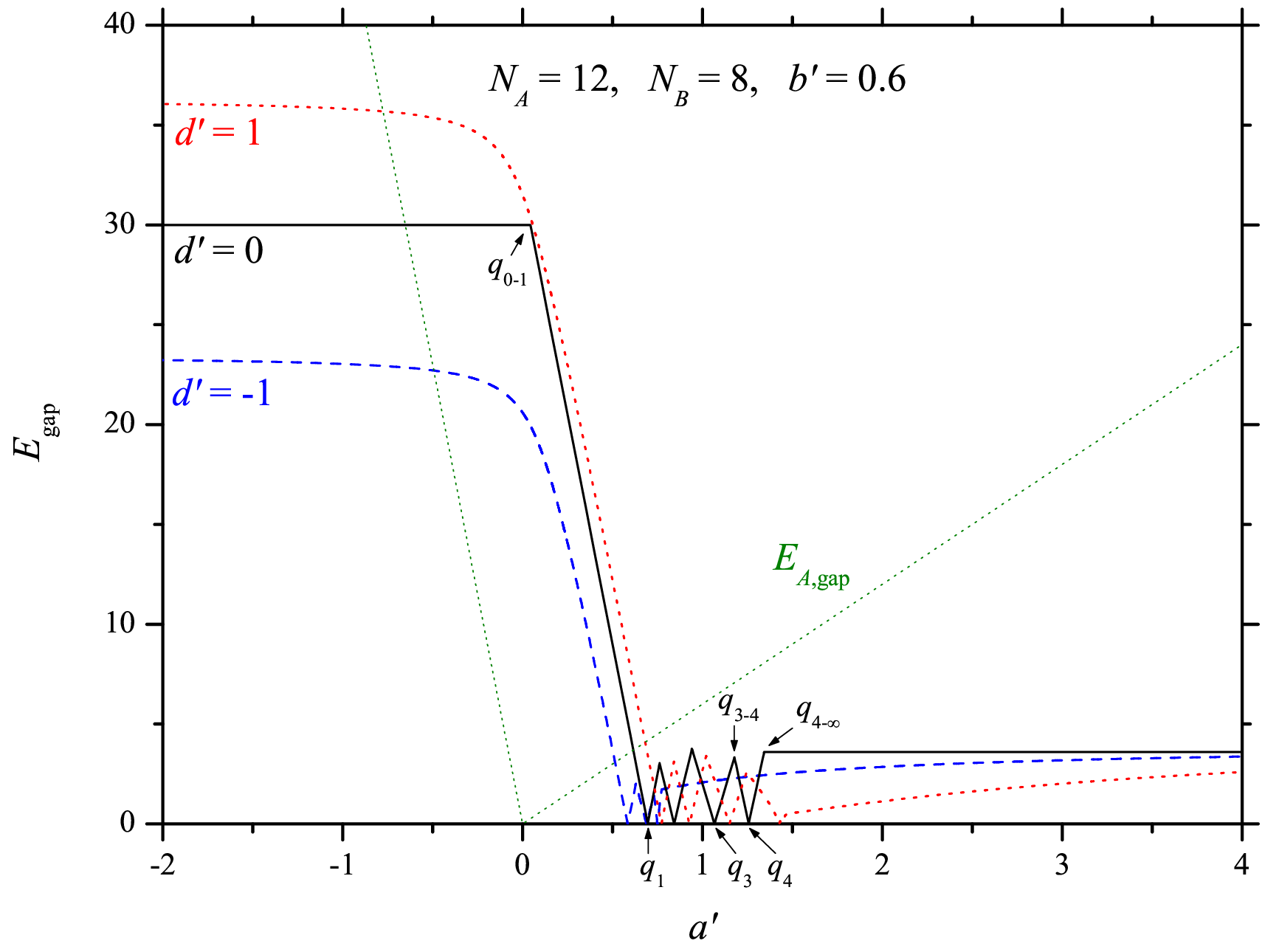}}
\caption{ (color online) The same as Fig.~\ref{fig2} but with $
b^{\prime }=0.6$. The critical points are marked by $
\cdots,q_i,q_{i-(i+1)},q_{i+1},\cdots$ }
\label{fig3}
\end{figure}

\begin{figure}[tbp]
\centering\resizebox{0.9\columnwidth}{!}{\includegraphics{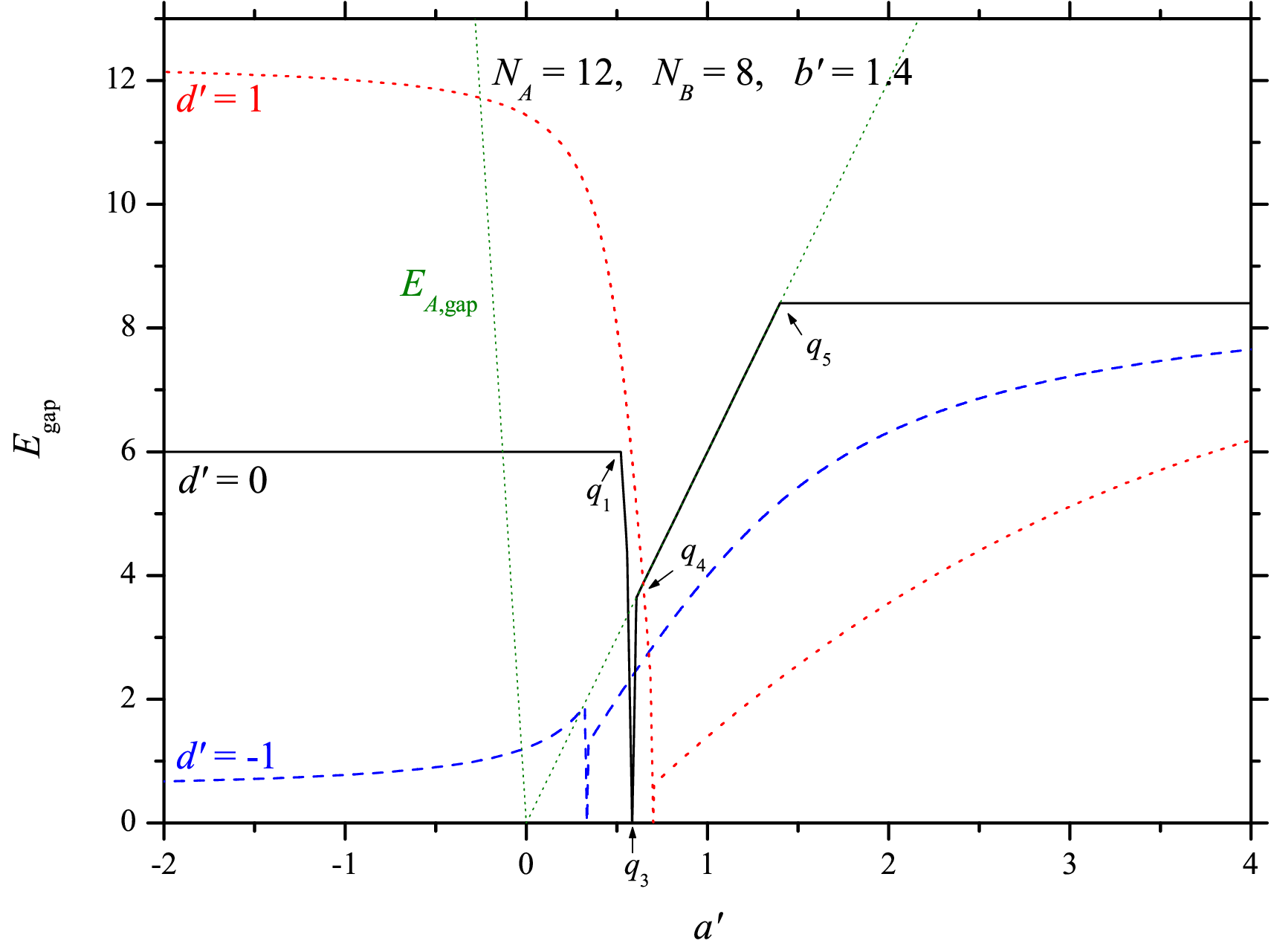}}
\caption{ (color online) The same as Fig.~\ref{fig2} but with $
b^{\prime }=1.4$. }
\label{fig4}
\end{figure}

With a comparison among the above modes, how $E_{\text{gap}}$ would vary
with $a^{\prime }$ when $b^{\prime }$ is fixed has been clarified as plotted
in Fig.~\ref{fig2} to Fig.~\ref{fig4} and the details are given in Table~\ref
{tab1} (with $b^{\prime }=-0.6$) and Table~\ref{tab2} (with $b^{\prime }=0.6$
and $1.4$). Let the gap of the $A$ -species alone be $E_{A,\text{gap}}$
which is also plotted in the figures for a comparison. In the figures and
tables, $a^{\prime }$ is divided into segments separated by the critical
points. E.g., we find a set of sawteeth in Fig.~\ref{fig2}. The $i$-th
sawtooth (counted from left to right) is situated in the segment $
(p_{i},p_{i+1})$ in which the g.s. is $(S_{A}^{i},N_{B})_{+}$, where $
S_{A}^{i}\equiv N_{A}-2i$. At $p_{i}$, the two textures $
(S_{A}^{i-1},N_{B})_{+}$ and $(S_{A}^{i},N_{B})_{+}$ are competing for the
g.s.. Therefore, from the equality $
E_{S_{A}^{i-1},N_{B}+}=E_{S_{A}^{i},N_{B}+}$ we know the location of $p_{i}$
as listed in the table. Besides, inside this segment, the two most
competitive modes for the lowest excitation are $E_{\text{x}S_{A}^{i}+}$ and
$E_{\text{x}S_{A}^{i}-}$. When $a^{\prime }$ is closer to $p_{i}$ ($p_{i+1}$
), the former (latter) is lower and \ going up (down) with $a^{\prime }$.
From the equality $E_{\text{x}S_{A}^{i}+}=E_{\text{x}S_{A}^{i}-}$ we can
find the critical point $p_{i-(i+1)}$ where the mode $E_{\text{x}S_{A}^{i}+}$
is replaced by $E_{\text{x}S_{A}^{i}-}$ leading to a peak emerging inside
the $i$-th sawtooth. Thus, when the two most competitive textures and the
two most competitive modes are found, the details as listed in the table can
be obtained.

\begin{table}[!htb]
\caption{ Details of the variation of the gap, as shown in Fig.~\ref{fig2}. In the first column, the critical point $p_i$  together with its next labeling a segment are listed.  $p_i$ ($p_{i+1}$) denotes the value of $a^{\prime }$ where the $i$-th sawtooth arises (ends), while $p_{i-(i+1)}$ denotes the value where the sawtooth is peaked. The analytical expression of $p_i$ is given in the second column, the associated numerical value arising from the parameters given for Fig.~\ref{fig2} is also given. The left-platform is ranged from $-\infty$ to $p_{0-1}$, then $
E_\text{gap}$ declines to zero at $p_1$ where the first sawtooth emerges. The
right-platform is ranged from $p_{N_A/2-\infty}$ to $\infty$. $E_\text{gap}$
increases linearly with $a^{\prime }$ from zero at $p_{N_A/2}$ to the
platform.
Then, the specification of the g.s., the mode of the lowest excitation, and
the associated $E_\text{gap}$ in each segment are successively listed in the third to fifth columns.  The index $i$ appearing in the third and fourth rows runs from $1$ to
$N_A/2-1$, with $S_A^i\equiv N_A-2i$. }
\label{tab1}
\begin{center}
\begin{tabular}{llllll}
\hline\hline
segment & $a^{\prime }$ & g.s. & mode & $E_\text{gap}$ &  \\ \hline
$-\infty\rightarrow p_{0-1}$ & $-\infty$ & $(N_A,N_B)_{N_A+N_B}$ & $E_{\text{
x}S}$ & $2(N_A+N_B)=40$ &  \\
$p_{0-1}\rightarrow p_1$ & $\frac{N_B-N_A}{2N_A-1}=-0.174$ & $
(N_A,N_B)_{N_A+N_B}$ & $E_{\text{x}S_A-}$ & $-a^{\prime }(4N_A-2)+4N_B$ & $
p_i\rightarrow p_{i-(i+1)}$ \\
&  &  &  &  &  \\
$p_i\rightarrow p_{i-(i+1)}$ & $\frac{N_B}{S_A^i+3/2}=1.067$ (if $i=3$) & $
(N_A-2i,N_B)_+$ & $E_{\text{x}S_A+}$ & $a^{\prime }(4N_A-8i+6)-4N_B$ &  \\
$p_{i-(i+1)\rightarrow}p_{i+1}$ & $\frac{N_B}{S_A^i+1/2}=1.231$ (if $i=3$) &
$(N_A-2i,N_B)_+$ & $E_{\text{x}S_A-}$ & $-a^{\prime }(4N_A-8i-2)+4N_B$ & $
p_{N_A/2}\rightarrow p_{N_A/2-\infty}$ \\
&  &  &  &  &  \\
$p_{N_A/2}\rightarrow p_{N_A/2-\infty}$ & $\frac{2N_B}{3}=5.333$ & $
(0,N_B)_{N_B}$ & $E_{\text{x}S_A+}$ & $6a^{\prime }-4N_B$ &  \\
$p_{N_A/2-\infty}\rightarrow +\infty$ & $\frac{2b^{\prime }+4N_B(1-b^{\prime
})}{6}=8.333$ & $(0,N_B)_{N_B}$ & $E_{\text{x}S_B-}$ & $b^{\prime
}(2-4N_B)=18$ &  \\ \hline\hline
\end{tabular}
\end{center}
\end{table}

\begin{sidewaystable}
 \caption{
 The same as Table~\ref{tab1} but with $b'=0.6$, $1.4$.
 Where the index $i$ in the third row is ranged from 1 to $j-1$ and $j$ can be obtained from the inequality Eq.~(\ref{sbs}).
 In particular, when $b'=0.6$ and $N_B=8$ we have $j=3$.
 }\label{tab2}
 \begin{center}
 \begin{tabular}{lllllll}
 \hline\hline
 segment                               & $a'$                                                                        & $b'$  & g.s.                  & mode                                          & $E_\text{gap}$                     \\ \hline
 $-\infty\rightarrow q_{0-1}$          & $-\infty$                                                                   & $0.6$ & $(N_A,N_B)_{N_A+N_B}$ & $E_{\text{x}S_B-}$                            & $-b'(4N_B-2)+4N_A=30$              \\
 $q_{0-1}\rightarrow q_1$              & $\frac{4N_B(b'+1)-4N_A-2b'}{4N_A-2}=0.0435$                                 & $0.6$ & $(N_A,N_B)_{N_A+N_B}$ & $E_{\text{x}S_A-}$                            & $-a'(4N_A-2)+4N_B$                 \\ \\
 $q_i\rightarrow q_{i-(i+1)}$          & $\frac{N_B}{(N_A-2i)+3/2-}=0.842$ (if $i=2$)                                & $0.6$ & $(N_A-2i,N_B)_+$      & $E_{\text{x}S_A+}$                            & $a'(4N_A-8i+6)-4N_B$               \\
 $q_{i-(i+1)}\rightarrow q_{i+1}$      & $\frac{N_B}{(N_A-2i)+1/2-}=0.941$ (if $i=2$)                                & $0.6$ & $(N_A-2i,N_B)_+$      & $E_{\text{x}S_A-}$                            & $-a'(4N_A-8i-2)+4N_B$              \\ \\
 $q_j\rightarrow q_{j-(j+1)}$          & $\frac{N_B}{(N_A-2j)+3/2-}=1.067$ (if $j=3$)                                & $0.6$ & $(N_A-2j,N_B)_+$      & $E_{\text{x}S_A+}$                            & $a'(4N_A-8j+6)-4N_B$               \\
 $q_{j-(j+1)}\rightarrow q_{j+1}$      & $\frac{N_B[2(N_A-2j+2)-b'(N_B+1)]}{(N_A-2j+2)(N_A-2j+3)}=1.178$ (if $j=3$)  & $0.6$ & $(N_A-2j,N_B)_+$      & $E_{\text{x},\text{qf}\rightarrow\text{pp}-}$ & $-E_{N_A-2j,N_B,+}$                \\
 $q_{j+1}\rightarrow q_{(j+1)-\infty}$ & $\frac{N_B[2(N_A-2j)-b'(N_B+1)]}{(N_A-2j)(N_A-2j+1)}=1.257$ (if $j=3$)      & $0.6$ & $(0,0)_+$             & $E_{\text{x},\text{pp}\rightarrow\text{qf}}$  & $E_{N_A-2j,N_B,+}$                 \\
 $q_{(j+1)-\infty}\rightarrow\infty$   & $\frac{2(N_A-2j)N_B+b'(6-N_B(N_B+1))}{(N_A-2j)(N_A-2j+1)}=1.343$ (if $j=3$) & $0.6$ & $(0,0)_+$             & $E_{\text{x},S_B+}$                           & $6b'=3.6$                          \\ \\
 $-\infty\rightarrow q_1$              & $-\infty$                                                                   & $1.4$ & $(N_A,N_B)_+$         & $E_{\text{x},S_B-}$                           & $4(N_A-b'N_B)+2b'=6$               \\
 $q_1\rightarrow q_2$                  & $\frac{N_B-2}{N_A-1/2}=0.522$                                               & $1.4$ & $(N_A,N_B)_+$         & $E_{\text{x},S_A-,S_B-}$                      & $a'(-4N_A+2)+b'(-4N_B+2)$          \\
                                       &                                                                             & $1.4$ &                       &                                               & $\ \ +4(N_A+N_B-2)=-46a'+30$       \\
 $q_2\rightarrow q_3$                  & $\frac{(N_B-2)[2(N_A-2)-b'(N_B-1)]}{(N_A-2)(N_A-1)}=0.557$                  & $1.4$ & $(N_A,N_B)_+$         & f-f$\rightarrow$p-p (collective mode)         & $-E_{N_A,N_B,N_A+N_B}=-123a'+91.2$ \\
 $q_3\rightarrow q_4$                  & $\frac{2N_AN_B-b'N_B(N_B+1)}{N_A(N_A+1)}=0.585$                             & $1.4$ & $(0,0)_+$             & p-p$\rightarrow$f-f (collective mode)         & $E_{N_A,N_B,N_A+N_B}=123a'-91.2$   \\
 $q_4\rightarrow q_5$                  & $\frac{2N_AN_B-b'N_B(N_B+1)}{N_A(N_A+1)-6}=0.608$                           & $1.4$ & $(0,0)_+$             & $E_{\text{x},S_A+}$                           & $E_{2,0,2}=6a'$                    \\
 $q_5\rightarrow+\infty$               & $b'=1.4$                                                                    & $1.4$ & $(0,0)_+$             & $E_{\text{x},S_B+}$                           & $E_{0,2,2}=6b'=8.400$              \\ \hline\hline
 \end{tabular}
 \end{center}
\end{sidewaystable}

There are the following distinguished features.

(i) When $a^{\prime }$ is sufficiently negative (positive), the g.s. has all
$A$-atoms in the f-phase (p-phase). Where, $E_\text{gap}$ appears as a platform
and depends on $a^{\prime }$ no more (i.e., the $A$-atoms are no more
involved in the lowest excitation).

(ii) There is a middle region with a set of sawteeth, in each sawtooth $\Psi
_\text{gs}=(S_{A}^{i},N_{B})_{+}$ (i.e., in a q-f phase) as mentioned above. This
middle region describes a process in which the increase of $a^{\prime }$
leads to the emergence of the 0-$A$-pairs pair-by-pair (or a decrease of $
S_{A}^{i}$ step-by-step) so that the $A$-species undergoes a series of $
(2\rightarrow 0)_{A}$ transitions. When $b^{\prime }$ is not so positive,
the decrease of $S_{A}$ is from $N_{A}$ to zero. However, when $b^{\prime }$
is sufficiently positive, due to the appearing of a low-lying collective mode
$E_{\text{x},\text{qf}\rightarrow \text{pp}}^{j}$ , a sudden jump from a
q-phase to a p-phase is possible. It implies that the total number of
sawteeth is reduced and the middle region is shortened (comparing Fig.~\ref{fig3} with
Fig.~\ref{fig2}).

(iii) There is a very special region with $0<a^{\prime }<p_{1}$ (the
left-down corner of the first quadrant in Fig.~\ref{fig1} where the g.s. is in the f-f
phase). In this region, when the two species are not mixed up, the g.s. is
in the p-p phase. However, an input of a few $B$-atoms causes the great
change from p-p to f-f. In this change, once a few $B$-atoms form a cluster,
the interspecies interactions from these $B-$atoms are mutually promoted
(because their spins are parallel) and would push the surrounding $A$-atoms
to allign along the same direction (if $c<0$) with them so as to form an $A$
-cluster. In return, the latter would push more $B$-atoms to align
resulting in a larger $B$-cluster, which is more powerful to break the 0-$A$
-pairs to enlarge the $A$-cluster, and vice, versa. This is a mechanism with
a positive feedback underlying the p-p$\rightarrow $f-f collective
transition. \cite{hlb25} Thus, while the inter-species interaction leads to
the polarized cluster, the intra-species interactions tend to keep the
0-pairs. There is an inter-intra competition, when $|c|$ is strong enough
(i.e., $a^{\prime }$ and $b^{\prime }$ are weak enough) the former wins and
leads to the f-f phase.

(iv) Recalled that, for single-species systems, the energy for a collective
excitation is usually high. However, for multi-species systems, the
collective excitation (say, q-f$\rightarrow $p-p) might be very low. Let $
S_{A}^{j}\equiv N_{A}-2j$, where $j$ is an integer to specify a specific
q-phase. The energy associated with the q-f$\rightarrow $p-p excitation is
\begin{equation}
E_{\text{x},\text{qf}\rightarrow \text{pp}}^{j}=-[a^{\prime
}S_{A}^{j}(S_{A}^{j}+1)+b^{\prime }N_{B}(N_{B}+1)-2S_{A}^{j}N_{B}],
\end{equation}
which could be low if $a^{\prime }$, $b^{\prime }$, and $j$ are suitably
chosen. This mode has to compete with the $E_{\text{x}S_{A}^{j}+}$ and $E_{
\text{x}S_{A}^{j}-}$ modes. One can prove that, if the inequality
\begin{equation}
\frac{S_{A}^{j}(S_{A}^{j}-2)}{(S_{A}^{j}-1/2)}<b^{\prime }(N_{B}+1)<\frac{
S_{A}^{j}(S_{A}^{j}+2)}{(S_{A}^{j}+3/2)}  \label{sbs}
\end{equation}
is satisfied, then $E_{\text{x},\text{qf}\rightarrow \text{pp}}^{j}<E_{\text{
x}S_{A}^{j}+}$ and $E_{\text{x},\text{qf}\rightarrow \text{pp}}^{j}<E_{\text{
x}S_{A}^{j}-}$ both hold, and the collective excitation q-f$\rightarrow $p-p
would replace the other two modes and the middle region ends as shown in
Fig.~\ref{fig3} and Table~\ref{tab2} (where $j=3$). Thus the above inequality can be used to
determine $j$, namely, the total number of the sawteeth.

\subsection{The eigenstates of $\hat{H}_{spin}$}

When the odd channel has been taken into account, making use of the
fractional parentage coefficients, diagonalization of the whole Hamiltonian
can be realized to obtain exact numerical solutions. The results for $E_{
\text{gap}}$ with $d^{\prime }=\pm 1$ are also plotted in Fig.~\ref{fig2} to
Fig.~\ref{fig4} We found that the qualitative features (the emergence of the
platforms, the sawtooth, and a number of gapless points) remain unchanged.
Nonetheless, there are quantitative modifications. In general, when $
d^{\prime }$ is negative (positive), the spin textures of the g.s. would be
modified such that the probability of two different bosons coupled to $
\lambda =1$, $P_{\lambda =1}$, would be enlarged (reduced). Besides, among
the competing candidates for the first excited state, the one leading to a
larger (smaller) $P_{\lambda =1}$ would be more favorable. For a simple
example, we study what happens in the spin-states $\psi _{1}\equiv \lbrack
(\chi (1)\chi (2))_{0}\chi (3)]_{1}$, $\psi _{2}\equiv \lbrack (\chi (1)\chi
(2))_{2}\chi (3)]_{3}$ and $\psi _{3}\equiv \lbrack (\chi (1)\chi
(2))_{2}\chi (3)]_{1}$ of a 3-body system where 1 and 2 denote two $A$-atoms
and 3 denotes a $B$-atom, $\chi $ is the spin state of a spin-1 atom. Then,
we found the probability of an $A$-atom and a $B$-atom coupled to $\lambda
=1 $ in the $\psi _{1}$ state is $\langle \psi _{1}|P_{\lambda =1}|\psi
_{1}\rangle =3(W(1,1,1,1,0,1))^{2}=\frac{1}{3}$ but $\langle \psi
_{2}|P_{\lambda =1}|\psi _{2}\rangle =0$, and $\langle \psi _{3}|P_{\lambda
=1}|\psi _{3}\rangle =0.417$. It implies that, when all the spins are
parallel as in $\psi _{2}$, $P_{\lambda =1}=0$, while the probability is
notable when a $B$-atom meets a 0-$A$-pair or meets a 2-$A$-pairs with a
reverse spin (as in $\psi _{3}$). However, $\psi _{3}$ is not a good choice
when $c<0$. Thus, in general, a negative $d^{\prime }$ would not favor the
formation of clusters, but favor the formation of 0-pairs (as $\psi _{1}$).
This explains why, in Fig.~\ref{fig2}, the point $p_{N_{A}/2}$ shifts to the
left when $d^{\prime }=-1$ (the p-$A$-phase can be realized when $a^{\prime
} $ is not so strong). Whereas in the curve with $d^{\prime }=+1$, $
p_{N_{A}/2} $ shifts to the right. A detailed study of the effect of the odd
channel is given in \cite{preprint}.

\section{The effect of impurity on medium-body systems}

Let the $B$-species be considered as an impurity ($N_{B}\ll N_{A}$). Based
on the above analytical formalism, the effect of impurity on the
spin-texture and on the stability of the g.s. can be studied in an
analytical way.

(i) A sudden and complete polarization

Due to the finding of the special region with $0<a^{\prime }<p_{1}\equiv
\frac{N_{B}}{N_{A}-1/2}$ and $0<b^{\prime }<q_{1}\equiv \frac{N_{A}}{
N_{B}-1/2}$ as shown in Fig.~\ref{fig1}, the magnetization of an anti-ferromagnetic $A$
-species in its g.s. can be maximized simply by adding impurity (with a
suitable $b^{\prime }$ so that the point $(a^{\prime },b^{\prime })$ is
located inside the domain where the g.s. is in the f-f phase). Let $\gamma
\equiv N_{B}/N_{A}$. When $N_{A}$ is not very small, the above first
inequality can be written as $0<a<\gamma |c|$. Thus, by choosing suitable
kind and number of $B$-atoms, the magnetization can be maximized.

(ii) A partial magnetization

Due to the finding of the q-f region as shown in Fig.~\ref{fig1}, the spin $S_{A}$ of
the g.s. of an anti-ferromagnetic species can be tuned to a specific value $
S_{A}^{i}\equiv N_{A}-2i$. This can be realized if the impurity is so chosen
to enable $a^{\prime }$ and $b^{\prime }$\ to be located inside a specific
q-f region with $S_{A}=S_{A}^{i}$. From Table~\ref{tab1} we know that this region is
bound as $p_{i}\leq a^{\prime }<p_{i+1}$. together with the condition $
E_{S_{A}^{i},N_{B,}+}<0$\ (if $E_{S_{A}^{i},N_{B,}+}>0$, the p-p phase would
replace the q-f phase). These conditions can be rewritten as
\begin{equation}
 \frac{N_{B}}{S_{A}^{i}+3/2}\leq a^{\prime }<\frac{N_{B}}{S_{A}^{i}-1/2}\label{qfa}
\end{equation}
and
\begin{equation}
b^{\prime }<\frac{S_{A}^{i}(2N_{B}-a^{\prime }(S_{A}^{i}+1))}{N_{B}(N_{B}+1)}\label{qfb}
\end{equation}
By choosing the kind (to choose $b$\ and $c$) and the number of $B$-atoms to
satisfy the above conditions, then the desired $S_{A}^{i}$, and accordingly
a desired magnetization can be achieved.

(iii) The stability of the g.s. (measured by the gap)

Note that, when the $A$-species is alone, the gap is $E_{A,\text{gap}
}=|a^{\prime }|(4N_{A}-2)$ (if $a^{\prime }<0$) or $E_{A,\text{gap}
}=6a^{\prime }$ (if $a^{\prime }>0$). By comparing $E_{A,\text{gap}}$ and $
E_{\text{gap}}$ as shown in the figures we find that there are domains in
which $E_{\text{gap}}\lessgtr E_{A,gap}$. Therefore, we can choose suitable
impurity to enable $E_{\text{gap}}$ being given at a specified value. In
particular, for any index $i$ ranging from 1 to $N_{A}/2$, if the impurity
is so chosen to enable $a^{\prime }=\frac{N_{B}}{S_{A}^{i}+3/2}$ and $
b^{\prime }$ satisfies Eq.(\ref{qfb}), then $E_{gap}$
would become zero (because meanwhile the point $(a^{\prime },b^{\prime })$
is located at the left end of a sawtooth shown in Fig.~\ref{fig2}).

(iv) Collective excitation

In Fig.~\ref{fig1} we find that, in the first quadrant, there is a curve separating
the p-p phase with other phases. The neighborhood lying along this curve is
a region in which the collective excitation becomes the lowest. One can
choose suitable impurity to enable $a^{\prime }$ and $b^{\prime }$\ falling
into this region so that the texture of the g.s. will undergo a great change
via a tiny perturbation. However, it is mentioned that the analytical
formulae hold only approximately, they could serve as a guide and should be
modified via the exact solutions with taking the $\lambda =1$ channel into
account as shown below.

\section{The effect of impurity on many-body systems}

When $N_{A}$ becomes larger, we found that the $\lambda =1$ channel becomes
more important and henceforth the above analytical solutions become less
accurate. Hence, the evaluation of the effect of impurity must be based on
the exact solutions in which the eigenstates are given by $\Psi
_{S,i}=\sum_{S_{A}S_{B}}C_{_{S_{A}S_{B}}}^{S,i}\phi _{S_{A}S_{B}S}$, where
the coefficients are obtained numerically from diagonalizing the Hamiltonian
given in Eq.~(\ref{hm}). In these cases we can define the averages $
\overline{S_{A}}$ and $\overline{S_{B}}$ as $\langle \Psi _{S,1}|\hat{
S_{X}^{2}}|\Psi _{S,1}\rangle \equiv \overline{S_{X}}(\overline{S_{X}}+1)$
for the g.s.. Just as in \cite{preprint}, we find again that these averages
can be used to specify the g.s.. Obviously, $\overline{S_{X}}$ is
proportional to the averaged magnetization of the $X$-species.

\begin{figure}[tbp]
\centering\resizebox{0.9\columnwidth}{!}{\includegraphics{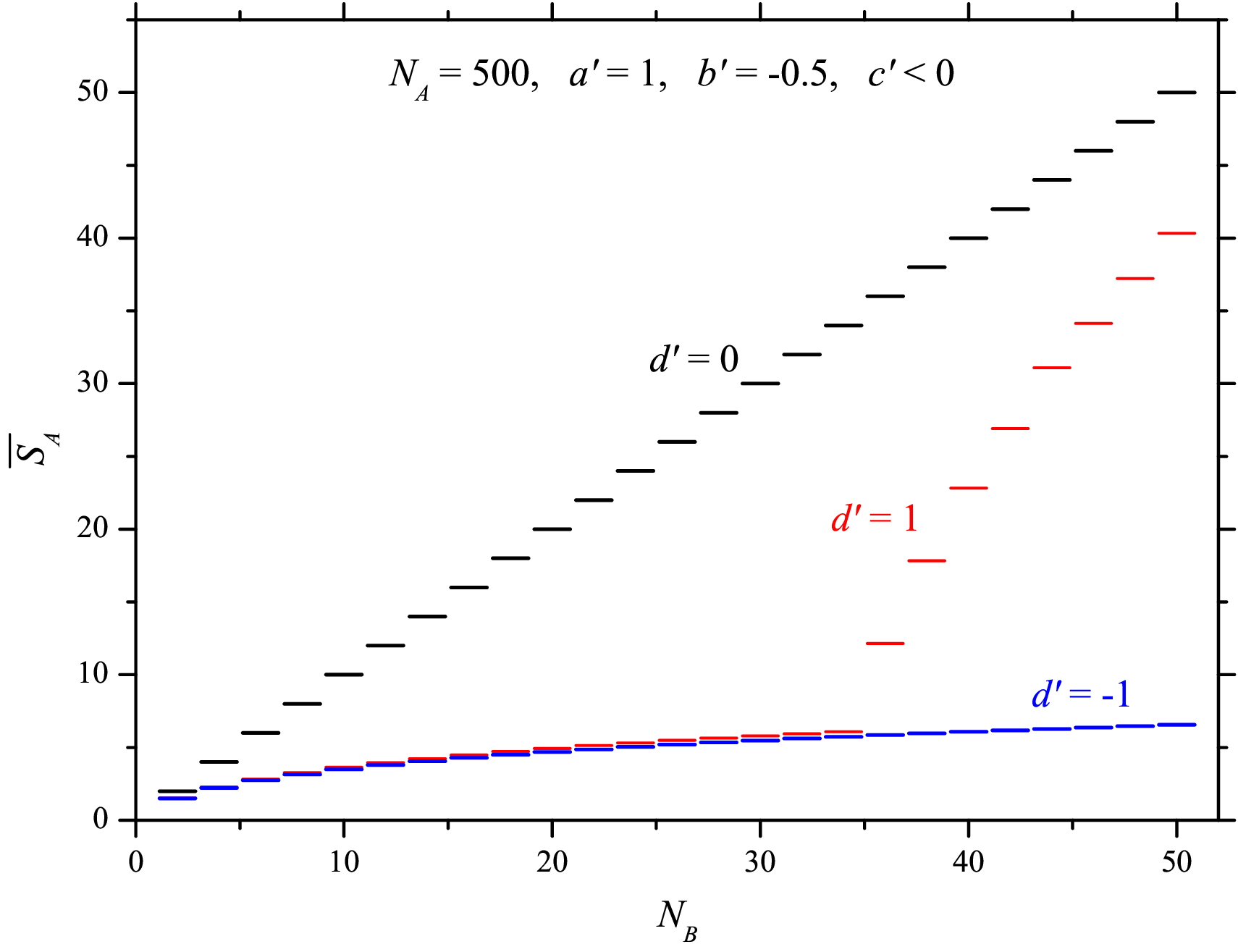}}
\caption{(color online) $\overline{S_A}$ (the averaged total spin of the $A$
-species) against $N_B$ (an even number). $N_A=500$, $a^{\prime }=1$, $b^{\prime }=-0.5$, $
d^{\prime }=0$, $\pm 1$, and $c^{\prime }<0$. }
\label{fig5}
\end{figure}

In the following we give an example for the case that $N_{A}$ is large ($
N_{A}=500$ and $N_{B}=2\rightarrow 50$ are assumed). Once the exact
numerical solutions are obtained, it is straight forward to obtain $
\overline{S_{A}}$. In Fig.~\ref{fig5} the $A$-atoms are anti-ferromagnetic while the $
B$-atoms are ferromagnetic, when $N_{A}$, $a^{\prime }$, $b^{\prime }$ and $
d^{\prime }$ are fixed, based on the exact solutions, the variation of $
\overline{S_{A}}$ against $N_{B}$ is plotted. Note that the $A$-atoms are
previously all in the 0-$A$-pairs. From the figure we know how the
magnetization can be induced by inputting certain number of $B$ -atoms.

In Fig.~\ref{fig5}, it is notable that there is a critical point lying at $N_{B}=34$
if $d^{\prime }=1$, and accordingly there are two types of competing spin
textures for the g.s.. The first type can be specified as $(\overline{S_{A}},
\overline{S_{B}})_{0}$ with $\overline{S_{A}}=\overline{S_{B}}\ll N_{B}$.
Thus it contains a small $B$-cluster and a small $A$-cluster, both containing the same number of particles, while all other particles are in the 0-$X$-pairs.
The particle number contained in
the clusters is fluctuating (not a constant) and the spins of the two
clusters are anti-parallel to enable $S=0$. Let the g.s. be denoted as $\Psi
_{\text{gs}}^{(d^{\prime },N_{B})}$. In the above case, since most particles
are in the 0-$X$-pairs while the two clusters are anti-parallel, the g.s. is
called in an anti-fluctuating p-p phase. With the set of parameters given
for Fig.~\ref{fig5} and with $d^{\prime }=1$.\ we have $\Psi _{\text{gs}
}^{(1,34)}=(6.080,6.080)_{0}$ (i.e., each cluster contains nearly six atoms).

The second texture can be specified as $(\overline{S_{A}},\overline{S_{B}}
)_{S}$ with $\overline{S_{A}}\gtrapprox \overline{S_{B}}<N_{B}$.\ It
contains two larger clusters with their spins being nearly parallel
resulting in $S\neq 0$ ($\frac{S}{N_{A}+N_{B}}$ remains small). Say, we have
$\Psi _{\text{gs}}^{(1,36)}=(12.152,11.284)_{20}$ which is called in a
parallel fluctuating q-q phase. Besides, when $2\leq N_{B}\leq 34$, all the
g.s. would be in the first type. When $N_{B}\geq 36$ the g.s. would be in
the second type. Say, we have $\Psi _{\text{gs}}^{(1,2)}=(1.540,1.540)_{0}$
and $\Psi _{\text{gs}}^{(1,50)}=(40.337,36.405)_{76}$. Thus the transition
at the critical point causes a sudden enlargement of the clusters together
with an alignment of their spins resulting in having a nonzero $S$. This
transition is caused simply by adding two more particles ($N_{B}$ from 34 to
36).

It is expected that the critical point of $N_{B}$ would vary with other
parameters. The details can be extracted from the exact numerical solutions
but will not be given here.

\begin{figure}[tbp]
\centering\resizebox{0.9\columnwidth}{!}{\includegraphics{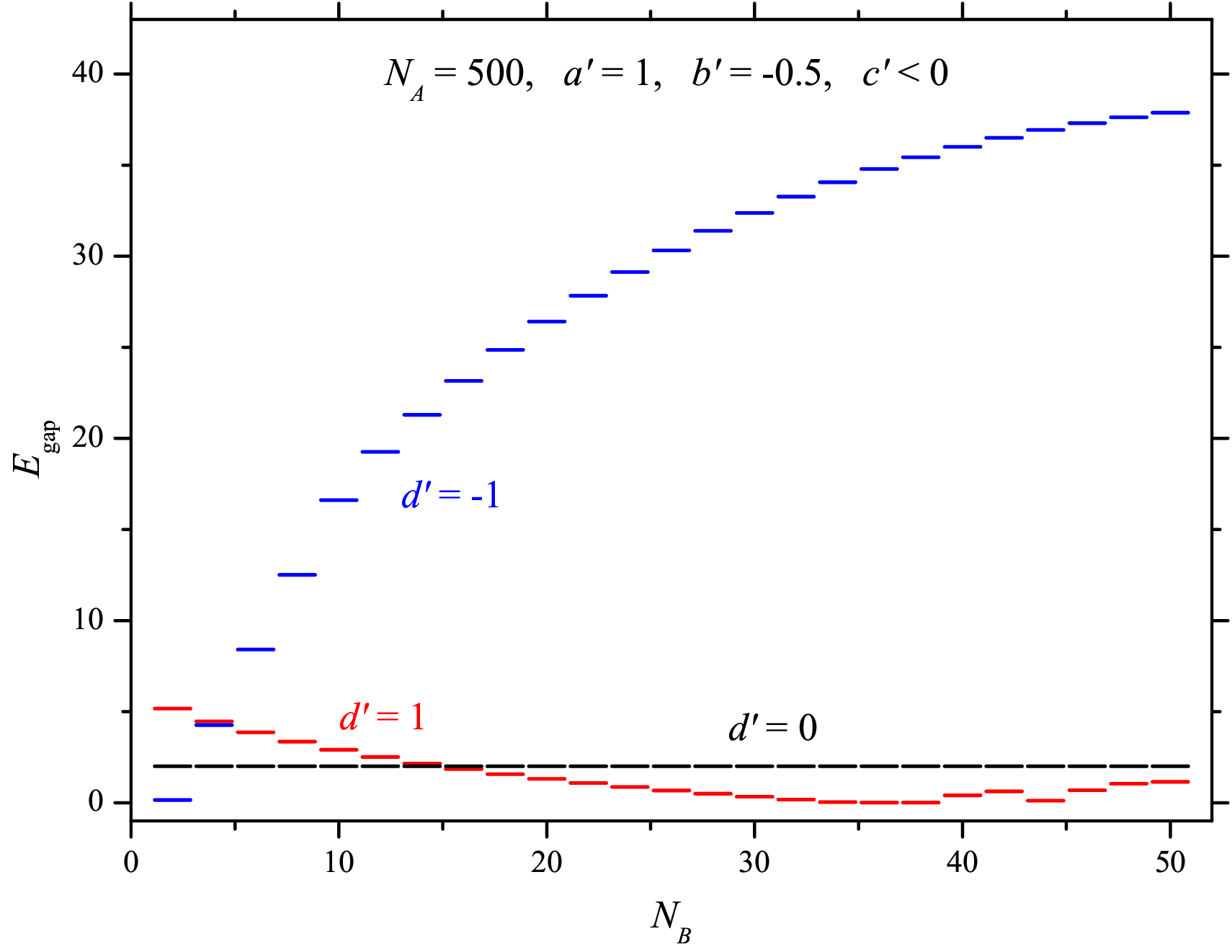}}
\caption{(color online) $E_\text{gap}$ against $N_B$. The other parameters
are the same as in Fig.~\ref{fig5}. }
\label{fig6}
\end{figure}

Fig.~\ref{fig6} is an example to show how the impurity number $N_{B}$ would
affect the stability of the g.s..

\section{Final remarks}

We have proposed an approach applicable for multi-species medium-body
ultracold systems with spin-1 atoms. Both exact numerical solutions and
approximate analytical solutions have been obtained. the latter is very
valuable for clarifying the underlying physics. The spin-texture of the g.s.
and various excitation modes have been studied. The effect of $c$
(disregarding being negative or positive) is to break the 0-pairs and to
form the polarized $A$-cluster and $B$-cluster, and will bring $\overline{
S_{A}}$ and $\overline{S_{B}}$ lying along the same direction if $c<0$ or
along two reverse directions if $c>0$. Whereas positive $a$ and/or positive $
b$ favor the formation of 0-pairs. The competition between the inter- and
intra-interactions is a distinguished feature of multi-species systems, it
underlies the variation of the spin textures.

For both cases $c>0$ and $c<0$, the results are qualitatively similar (the
main difference is the relative orientation of $\overline{S_{A}}$ and $
\overline{S_{B}}$). For instance, the series of f-f$\rightarrow $q-f$
\rightarrow $p-p transitions in the g.s., the platforms and the sawteeth in $
E_\text{gap}$, and the f-f$\rightarrow $p-p collective transition are all found
in both cases. However, the details with $c>0$ are not given here to avoid
being tedious.

It turns out that our approach can serve as a rigorous theory for studying
the effect of impurity which is important to the preparation of ultracold
artificial materials. When the $B$-atoms are considered as impurity and have
been input into anti-ferromagnetic $A$-atoms, the spin-texture and the
stability (measured by the gap) of the g.s. of the mixture can be changed
greatly. The appearance of the $B$-atoms as a polarized cluster provides a
specific direction; the attractions (if $c<0$) or repulsions (if $c>0$)
arising from each $B$-atom are mutually reinforced. While the 0-$A$-pair,
disregarding how many they are, are scarcely mutually affected (because the
effects from the two members of a pair are mutually canceled). In
particular, when $|c|$ and $N_{B}$ are appropriately chosen, without
imposing an external magnetic field, $S_{A}$ can be tuned to a given value
so that a desired magnetization vector can be realized. The energy gap can
also be tuned to a given value. Recall that the collective mode is usually
very high in single-species systems. However, this mode can be tuned to be
low-lying in multi-species systems. The g.s. can be even tuned to be
gapless. A number of formulae have been derived to guide the tuning.

Another distinguished feature of multi-species systems is the appearance of
the odd-channel. For medium-body systems, when the odd-channel has been
taken into account, the exact numerical solutions are found to be
qualitatively close to the approximate analytical solution. Thus, the
analytical solutions can be used not only to understand the underlying
physics but also to guide the selection for appropriate $B$-species to
change the properties of the $A$-species. Nonetheless, for many-body systems
($N_{A}$ is large), the exact solutions are found to deviate from the
analytical solutions explicitly. In this case the exact solutions alone are
sufficient to guide the way to change the $A$ -species. In particular, the
effect of the particle number of the impurity has been studied. This study
can be used to guide the tuning of $N_{B}$ for the preparation of artificial
material.

Ultimately, in the field of ultracold physics, the study of medium-body
systems appears to be worthy because they can be more accurately solved in
the theoretical aspect and can be more concisely controlled in practice.
Furthermore, since the effect of the odd-channel becomes large when the
particle number increases, for the cases with $d^{\prime }$ being not so
weak, the existing many-body theories should be revised if the odd channel
has not been taken into account.

\section*{Competing Interests}

The authors declare no competing interests.

\section*{Data Availability}

All data generated or analyzed during this study are included in this
published article.

\section*{Author Contributions}

Y. Z. He is responsible for the theoretical derivation and numerical
calculation. C. G. Bao conceived the idea, performed the theoretical
analysis, wrote the paper, and takes responsibility for the whole work.

\section*{Funding Declaration}

No funding.

\section*{Appendix: Fractional parentage coefficients of spin-1 systems}

We consider a spin-1 system containing $N$ particles of the same species
governed by the Hamiltonian $\sum_{\lambda,i>j}\tilde{g}_\lambda
P_\lambda^{i,j}$. The total spin $S$ and its $Z$-component $M$ are
conserved. Thus, the normalized and symmetrized eigenstates can be specified
by $S$ and $M$ as $\theta_{SM}^N$, where $N-S$ must be even; otherwise, $
\theta_{SM}^N$ is zero. It has been proved that the multiplicity of $
\theta_{SM}^N$ is one. \cite{katr} Thus, $\theta_{SM}^N$ is unique, and the
set $\{\theta_{SM}^N\}$ is complete. One can extract a particle, say, the $i$
-th particle, from this state. After the extraction, the total spin of the
rest part must be either $S+1$ or $S-1$. The other choice is not possible
due to the even-odd parity. Thus, we have
\begin{equation}
\theta_{SM}^N = a_S^N [ \chi(i)\theta_{S+1}^{N-1}]_{SM} +b_S^N
[\chi(i)\theta_{S-1}^{N-1}]_{SM},  \label{fpc1}
\end{equation}
where $\chi(i)$ is the spin state of the $i$-th particle.

These coefficients have been studied previously and, as proven in Ref.~\cite{fpc}, are
\begin{equation}
a_S^N = [\frac{(N-S)(S+1)}{N(2S+1)}]^{1/2}, \ \ \ b_S^N = [\frac{(N+S+1)S}{
N(2S+1)}]^{1/2}.
\end{equation}
They are called the 1-body fractional parentage coefficients.

For convenience, Eq.~(\ref{fpc1}) can be rewritten as
\begin{equation}
\theta_{SM}^N = \sum_{S^{\prime }} a_{S^{\prime }S}^N
[\chi(i)\theta_{S^{\prime }}^{N-1}]_{SM},  \label{fpc1p}
\end{equation}
where
\begin{equation}
a_{S^{\prime }S}^N = [\frac{(N-S)(S+1)}{N(2S+1)}]^{1/2} \delta_{S^{\prime
},S+1} +[\frac{(N+S+1)S}{N(2S+1)}]^{1/2} \delta_{S^{\prime },S-1}.
\end{equation}
When one more particle, say, the $j$-th particle, is further extracted from
Eq.~(\ref{fpc1}), we have
\begin{equation}
\theta_{SM}^N = \sum_{\lambda,S^{\prime }} h_{\lambda S^{\prime }S}^N
\{[\chi(i)\chi(j)]_\lambda\theta_{S^{\prime }}^{N-2}\}_{SM},  \label{fpc2}
\end{equation}
where $\chi(i)$ and $\chi(j)$ are coupled to $\lambda$. They both have been
extracted, and
\begin{eqnarray}
h_{0,S,S}^N &=& [\frac{(N-S)(N+S+1)} {3N(N-1)}]^{1/2}, \\
h_{2SS}^N &=& [\frac{S(2S+2)(N-S)(N+S+1)} {3(2S-1)(2S+3)N(N-1)}]^{1/2}, \\
h_{2,S+2,S}^N &=& [\frac{(S+1)(S+2)(N-S)(N-S-2)} {(2S+1)(2S+3)N(N-1)}]^{1/2},
\\
h_{2,S-2,S}^N &=& [\frac{S(S-1)(N+S+1)(N+S-1)} {(2S-1)(2S+1)N(N-1)}]^{1/2},
\end{eqnarray}
and $h_{\lambda S^{\prime }S}^N=0$, if $S^{\prime }$ does not belong to the
above cases.

The coefficients in Eq.~(\ref{fpc2}) are called 2-body fractional parentage
coefficients.

With the above formulae and coefficients, the matrix elements of any 1-body
and 2-body operators against the eigenstates can be conveniently and
analytically derived. This is a great advantage because all relevant
physical quantities of the system can be obtained without knowing the
details of $\theta_{SM}^N$.

\end{document}